\pdfoutput=1
\documentclass[twocolumn,showpacs,preprintnumbers,amsmath,amssymb,superscriptaddress]{revtex4-1}
\usepackage{epsfig}
\usepackage{graphicx}
\usepackage{dcolumn}
\usepackage{bm}
\usepackage{float}
\usepackage{color}

\begin{document}

\selectfont

\title{\color{black}Modeling the topology of protein interaction networks}

\author{Christian M. Schneider}
\email{schnechr@ethz.ch}
\affiliation{Computational Physics, IfB, ETH Zurich, Schafmattstrasse 6, CH-8093 Zurich, 
Switzerland}
\author{Lucilla de Arcangelis}
\affiliation{Computational Physics, IfB, ETH Zurich, Schafmattstrasse 6, CH-8093 Zurich, 
Switzerland}
\affiliation{Department of Information Engineering and CNISM, Second University of
Naples, I-81031 Aversa (CE), Italy}
\author{Hans J. Herrmann}
\affiliation{Computational Physics, IfB, ETH Zurich, Schafmattstrasse 6, CH-8093
Zurich, Switzerland}
\affiliation{Departamento de F\'{\i}sica, Universidade Federal do Cear\'a, 60451-970
Fortaleza, Cear\'a, Brazil}

\date{\today}

\begin{abstract}
A major issue in biology is the understanding of the interactions between proteins. These interactions can be described by a network, where the proteins are modeled by nodes and the interactions by edges. The origin of these protein networks is not well understood yet. Here we present a two-step model, which generates clusters with the same topological properties as networks for protein-protein interactions, namely, the same degree distribution, cluster size distribution, clustering coefficient and shortest path length. The biological and model networks are not scale free but exhibit small world features. The model allows the fitting of different biological systems by tuning a single parameter.

\end{abstract}

 \pacs{64.60.aq, 
       89.75.Fb, 
       87.15.km, 
       87.23.Kg 
       }
\keywords{network,robustness,topology,malicious attack}

\maketitle
\begin{figure*}
 \includegraphics[width=6.5cm,angle = 0]{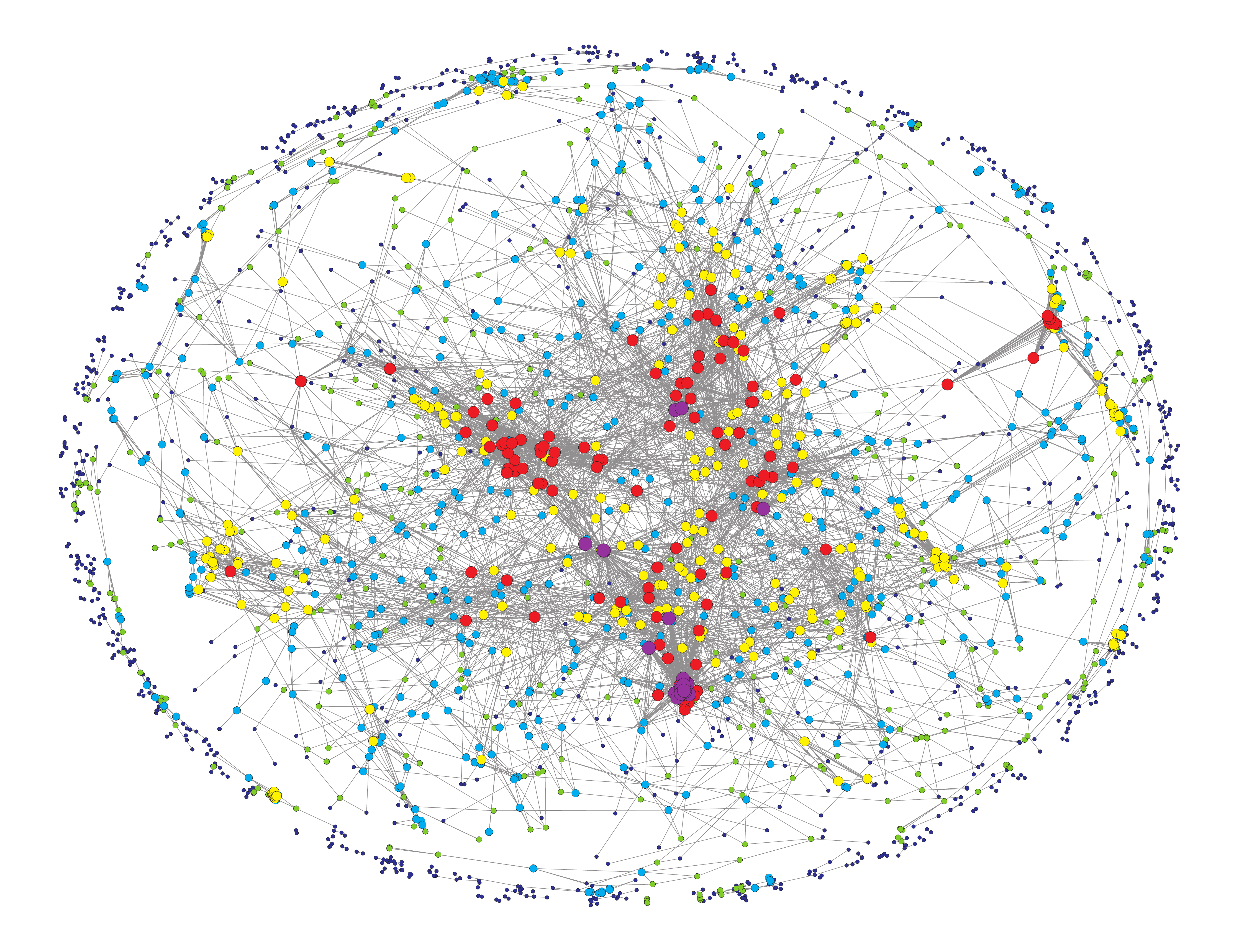}\hspace{2cm}
 \includegraphics[width=6.5cm,angle = 0]{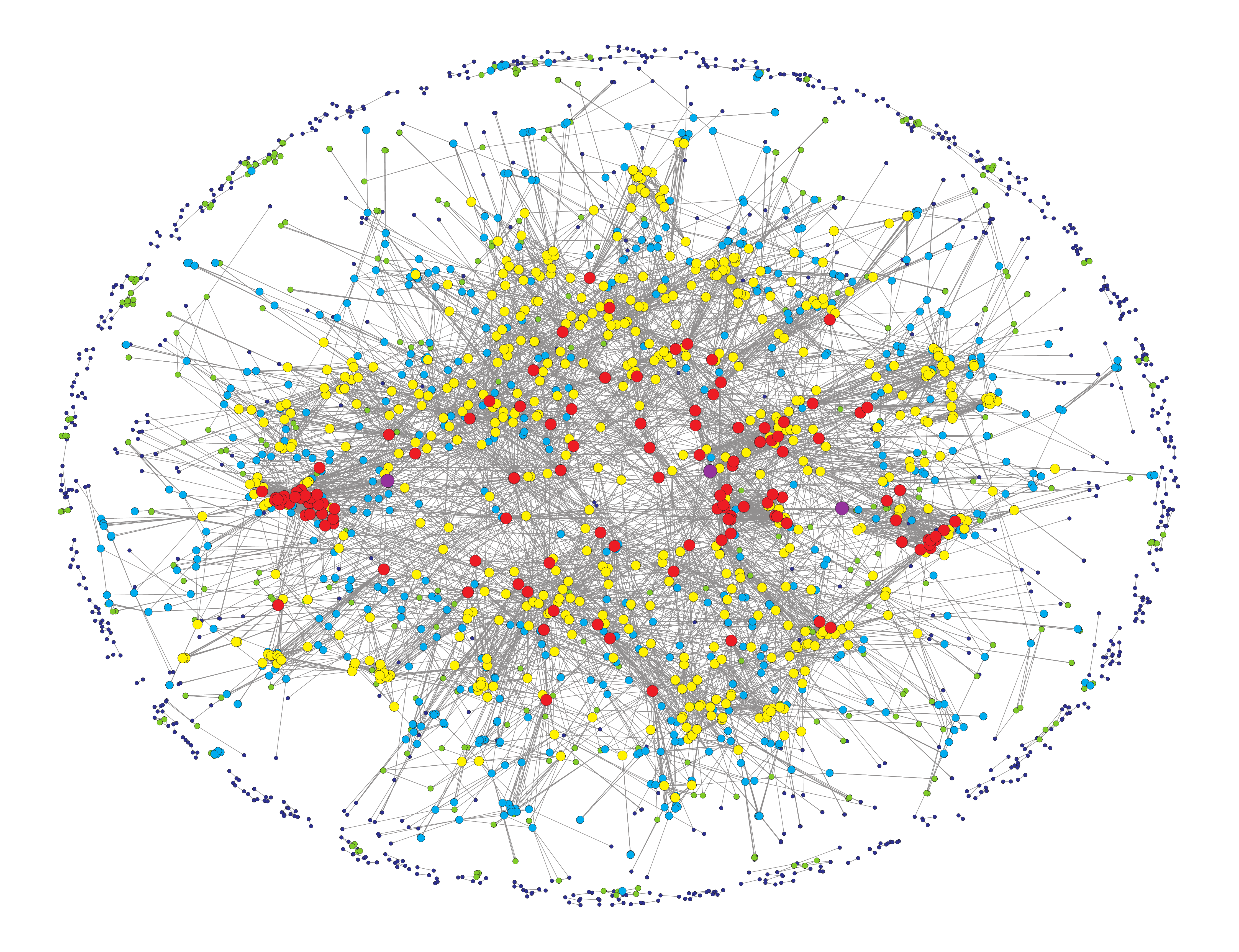}
 \caption{(Color online) The topology of (left) the AH network and (right) the model network with $\alpha = 0.75$. The largest clusters are drawn in the center and the smaller clusters on the border. The largest clusters are drawn in the center and the smaller clusters on the border. The color code and the size (from small to large) represent the degree of each site on a logarithmic scale: blue $k < 3$, green $3\le k < 5$, cyan $5\le k < 10$, yellow $10\le k < 21$, red $21\leq k < 43$ and purple $k\geq43$\cite{Pajek08}.}
\label{fig:1}
\end{figure*}
\section{Introduction}
The study of complex networks in biology promises new fruitful insights about the functionality of genes and proteins \cite{Hartwell99,Jeong00,Uetz00,Gavin02,barev}. Since the interactions between proteins determine their functionality, the properties and the origin of the interaction networks have attracted much attention \cite{Ideker01,Hood03,Editorial}. They consist of protein complexes, which are connected in a large, constantly evolving, cluster \cite{schwi}. The analysis of hundreds of protein complexes has established that some of the relevant structural features are the contact area, the shape of the interfaces, the complementarity of surface shapes, and the interaction-mediating forces. Although not all interactions have been discovered yet, numerous studies have been performed and many data sets are available \cite{Friedman,DeRisi,Spellman,Gu,Yeast,Tang,Hakes}. One important outcome of these studies is that most protein networks show a wide range of variability in the number of nodes and edges and the average connectivity degree (Table \ref{tab1}). They appear not to be scale-free, namely, the distribution of connectivity degrees is not a power law although it stretches over a significant number of orders of magnitude. Moreover, they do not consist of one single cluster but in addition to a large component many small clusters of interactions are also detected.\\
These results suggest that the specific features of biological networks express different underlying mechanisms than do other networks, like social interaction networks or the internet \cite{Internet,barabreview}. In fact, it has been speculated that gene duplication is the dominant evolutionary force in shaping biological networks \cite{Bhan,Friedman}. Conversely, non-biological networks are typically driven by additive growth processes \cite{barabreview} such as, for instance, preferential attachment \cite{ba}, but many other mechanisms like rewiring \cite{rewire}, aging \cite{aging}, or fitness \cite{fitness} have been investigated. However, none of these models can reproduce the full topology of protein networks like, for instance, the emergence of isolated clusters found in real biological networks (Fig. \ref{fig:1}).\\
\begin{table}
 \begin{tabular}{l|c|c|c|c}
 {\it Organism} & $N$ & $M$ & $\langle k \rangle$ & $\alpha$\\\hline
Nocardia farcinica (NF) & 3582 & 12045 & 6.7 & 1.50\\
Bradyrhizobium japonicum (BJ) & 4883 & 19261 & 7.9 & 1.00 \\
Aeromonas hydrophila (AH) & 2708 & 9050 & 6.7 & 0.75\\
Citrobacter koseri (CK) & 3373 & 8212 & 4.9 & 0.50\\
Escherichia coli (EC) & 3204 & 13091 & 8.2 & 0.75\\
Pseudomonas aeruginosa (PA)& 3794 & 14252 & 7.5 & 0.75\\
Serratia proteamaculans (SP)& 3373 & 8187 & 4.9 & 0.75\\
Vibrio cholerae (VC)& 2512 & 8612 & 6.9 & 1.00\\
{Saccharomyces Cerevisae (SC)} & 4771 & 54607 & 22.9 & 1.75\\
{Homo Sapiens (HS)} & 11102 & 136930 & 24.7 & 1.75\\
\end{tabular}
\caption{List of organisms from STRING 8.2 data set \cite{string} investigated here. Columns report the number of nodes $N$, the number of edges $M$, the average degree $\langle k \rangle$, and the value of the model parameter $\alpha$ used here. Edges between pairs of proteins represent an $80\%$ reliability of protein interaction. NF belongs to Acatinobacteria, BJ to Alphaproteobacteria, and all other bacteria belong to the Gammaproteobacteria class.}
\label{tab1}
\end{table}
{\color{black} Here we propose a different model, which reproduces many topology properties of protein interaction networks. We do not consider the details of the biochemical mechanisms at the basis of each interaction, nor classify proteins in classes as in other approaches \cite{Uetz00,schwi,PNAS98}. Conversely, we follow a simple probabilistic approach.
\section{The Model}
The procedure starts with a fully connected network of $N$ sites and $M = N(N - 1)/2$ edges. The number of nodes is equal to the number of nodes of the biological network considered, $N = N_\text{bio}$. The evolution is performed according to the following steps:\\
(i) Choose at random a node $i$.\\
(ii) Choose at random an edge $e_{ij}$ and remove it with a probability $p_{i,j}$ related to the degree $k_j$ of the neighbor $j$ of node $i$:
\begin{eqnarray}
 p_{i,j} = \frac{p_j}{N_i}~~\text{with}~~p_j = \begin{cases} k_j^{-\alpha} & k_j > 1
\\ 0 & \text{otherwise} \end{cases}
\end{eqnarray}
and with $N_i$ the normalization $N_i = \sum_{l=1}^{k_i}{p_l}$. $\alpha>0$ is the only free parameter of the model and controls the relative robustness of edges belonging to highly connected nodes with respect to edges of sites with low $k$. This rule implies that ``the poor get poorer´´. The case $\alpha=0$ implies that all sites have the same probability to lose edges and the process reduces to a random depletion.\\
(iii) Repeat this procedure for another node $i$ until the number of edges $M$ in the network equals the number of nodes $N$.\\
(iv) Choose at random two nodes $i$ and $j$. Add an edge between these nodes with probability
\begin{eqnarray}
 p_{i,j} = \lbrack N_c(i,j) \rbrack^2/(k_i k_j),
\end{eqnarray}
where $N_c(i,j)$ is the number of neighbors that nodes $i$ and $j$ have in common. This step supposes that, if two given nodes are able to interact with the same nodes, they have a high probability to interact with each other.\\
(v) Repeat this procedure for another random pair of nodes $i$ and $j$ until the number of edges $M$ in the network equals the number of edges of the modeled biological network $M_\text{bio}$.\\
These rules are based on the assumption that the evolution is controlled by two basic mechanisms: (i) preferential depletion: the lower the node degree, the lower the probability to maintain interactions \cite{Schneider11}; (ii) similarity: the more common neighbors two nodes share, the higher is the probability to have an interaction.\\
The first mechanism is important for the emergence of isolated clusters and a maximal degree, while the second one is necessary to generate networks with a high clustering coefficient and assortativity. It is interesting to notice that the implementation of the depletion mechanism alone generates scale free networks and does not reproduce the topology of protein-protein interacion networks \cite{Schneider11}.}\\
\begin{figure}
 \includegraphics[width=4.8cm,angle = -90]{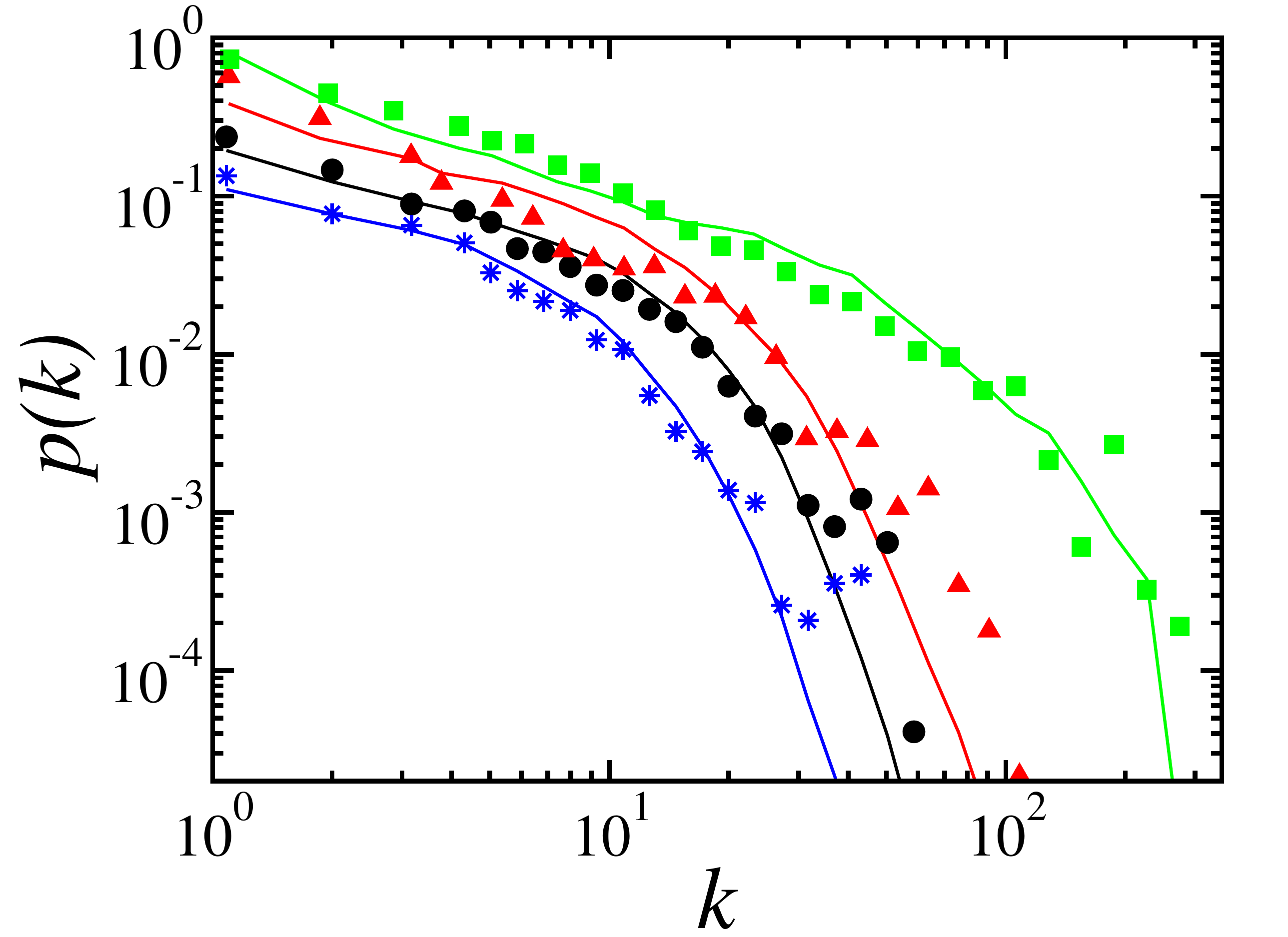}
 \caption{(Color online) The degree distribution $p(k)$ for AH (circles), BJ (triangles), CK (stars) and HS (squares) and their corresponding model networks (lines) with $\alpha$ obtained from Table \ref{tab1}. Star, triangle, and square data sets are shifted vertically by factors of $0.5$, $2$, and $5$, respectively, for better visibility.}
\label{fig:2}
\end{figure}
\section{Results}
\begin{figure}
 \includegraphics[width=4.8cm,angle = -90]{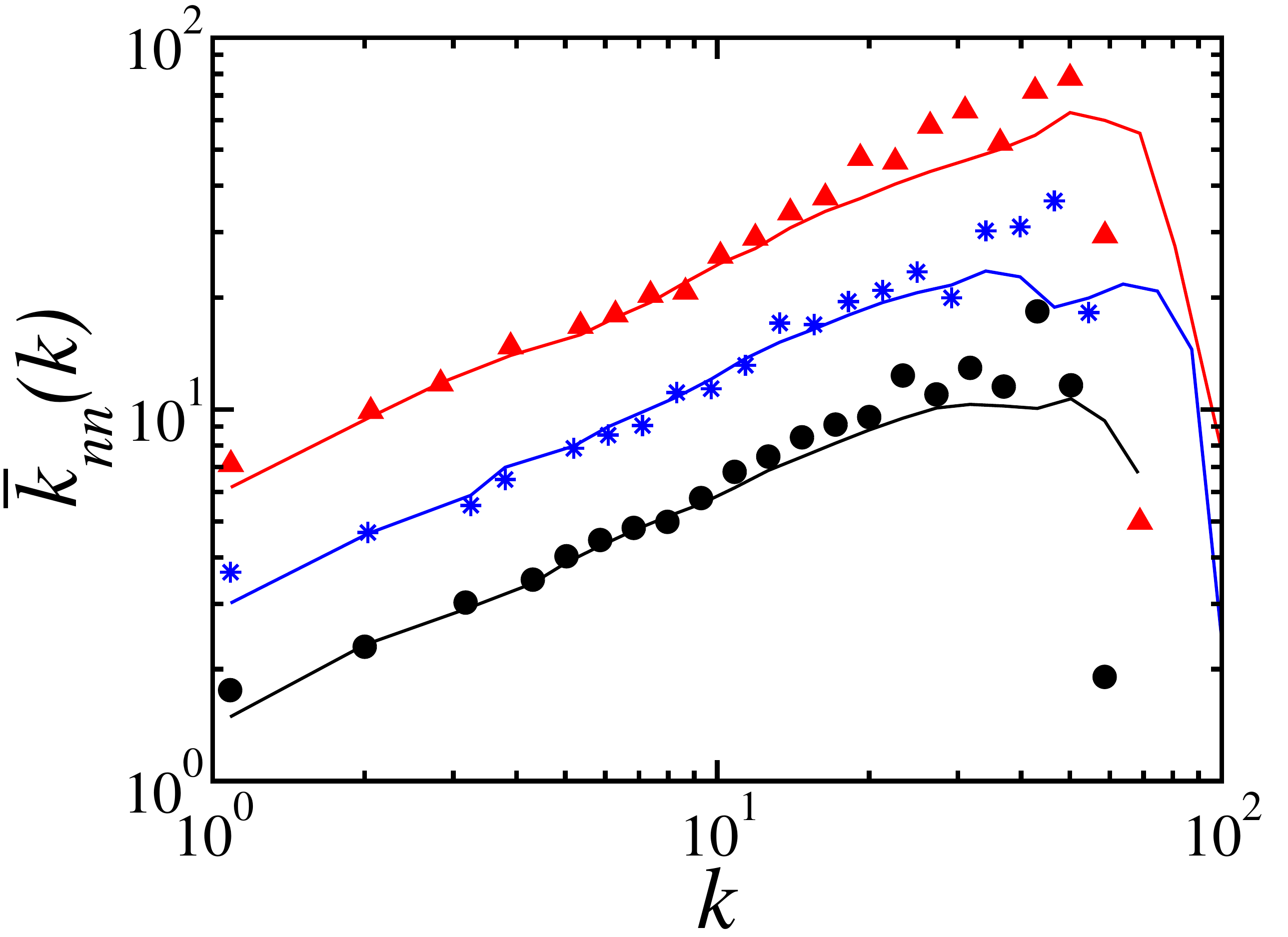}
 \caption{(Color online) Average nearest neighbor degree $\overline{k}_{nn}(k)$ of nodes of degree $k$ versus $k$ for AH (circles), PA (triangles) and VC (stars) and their corresponding model networks (lines). Top and bottom data sets are shifted vertically by a factor of $2$ and $0.5$.}
\label{fig:6}
\end{figure}

The biological networks are obtained from the STRING 8.2 data set \cite{string}, where a combined score of $80\%$ is used to decide whether two proteins interact. We tested our algorithm on the ten different biological networks listed in Table \ref{tab1}. {\color{black}For each organism we determine a value of the parameter $\alpha$ which provides a good fit (Table \ref{tab1}) for the degree distribution. All results for model networks are averages over $100$ independent runs for bacteria and $10$ runs for the other two networks.} In Fig. \ref{fig:1} we show an example for a biological network and the corresponding model network, with the same number of nodes and edges and $\alpha = 0.75$. Both networks have one large cluster with dangling ends, shown in the center of both graphs. Moreover, both networks have a large number of small clusters, placed on the border of each network. For both networks highly connected nodes are placed in the largest cluster, whereas small clusters are made of low-degree nodes. Since the topology is not a quantified differentiation property to decide whether two networks are similar, we calculate some fundamental properties characterizing the connectivity and the structure of the two networks. The model has by construction the same numbers of nodes $N$ and of edges $M$ as the biological one and therefore the average degrees per node $\langle k \rangle$ are exactly the same. To provide more information on the connectivity level of the two networks, we measure first the degree distribution. In Fig. \ref{fig:2} we show the degree distribution of different biological networks  and their numerical counterparts. The biological networks are not scale-free and the numerical data reproduce the data very well by tuning the parameter $\alpha$. We observe that the value of the exponent $\alpha$ controls the maximum degree and the exponential cutoff of the distribution. For $\alpha = 0$ the exponential cutoff is at $k = 1$ and therefore the degree distribution a pure exponential. By increasing $\alpha$, the range of the initial regime increases and the exponential cutoff moves toward larger $k$ values. {\color{black} To tune the parameter, we compare the tail of the degree distribution for different $\alpha$ values and choose the one which fits best.}
\begin{figure}
 \includegraphics[width=4.8cm,angle = -90]{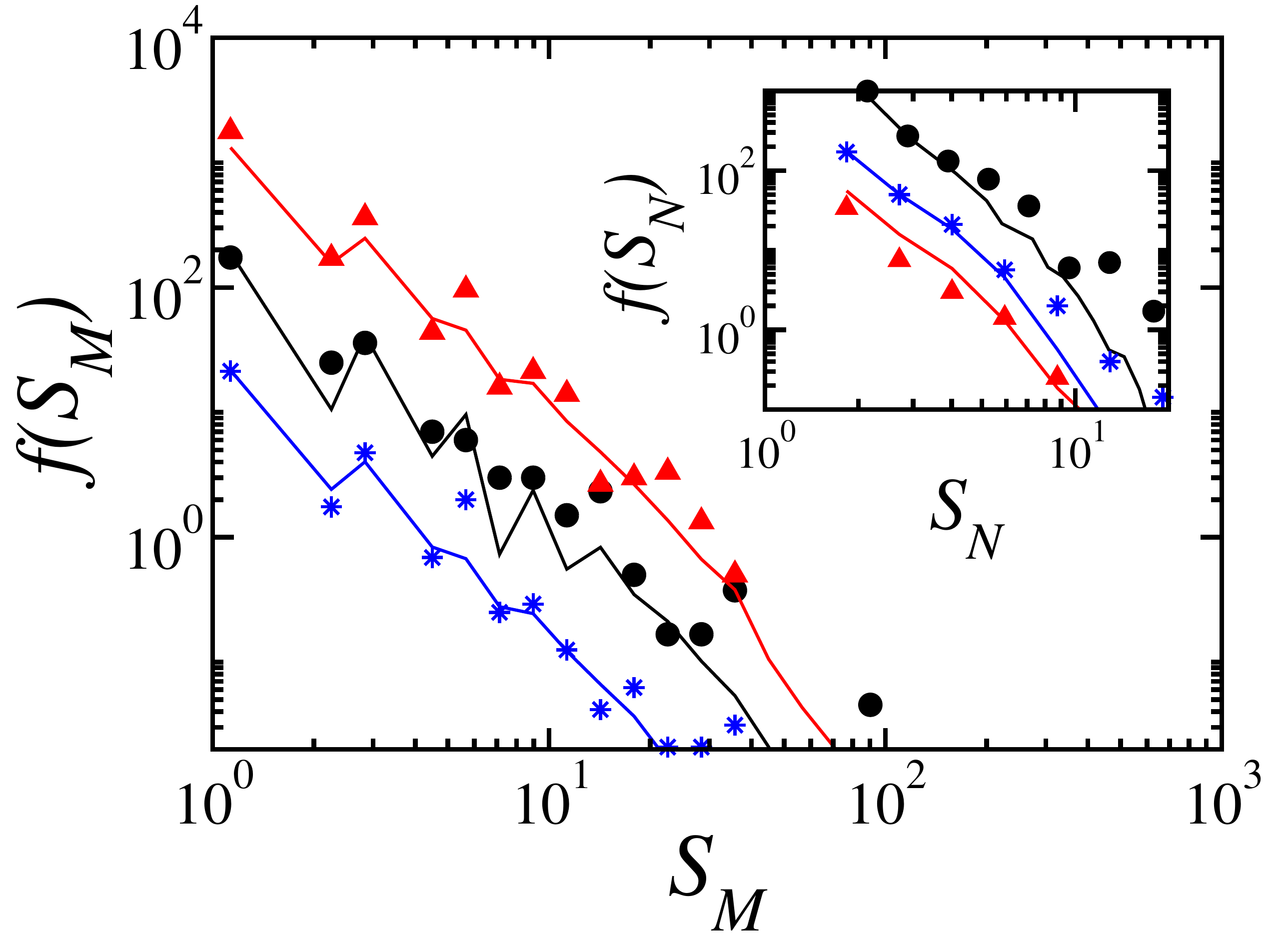}
 \caption{(Color online) Frequency $f(S)$ of finding a cluster with a given number of nodes $S_N$ for CK (circles), EC (triangles), and VC (stars) and with a given number of edges $S_M$ (inset) for AH (circles), BJ (triangles), and VC (stars), and their corresponding model networks (lines). Top and bottom data sets are shifted vertically by one decade, upward and downward.}
\label{fig:3}
\end{figure}
\begin{figure}
 \includegraphics[width=4.8cm,angle = -90]{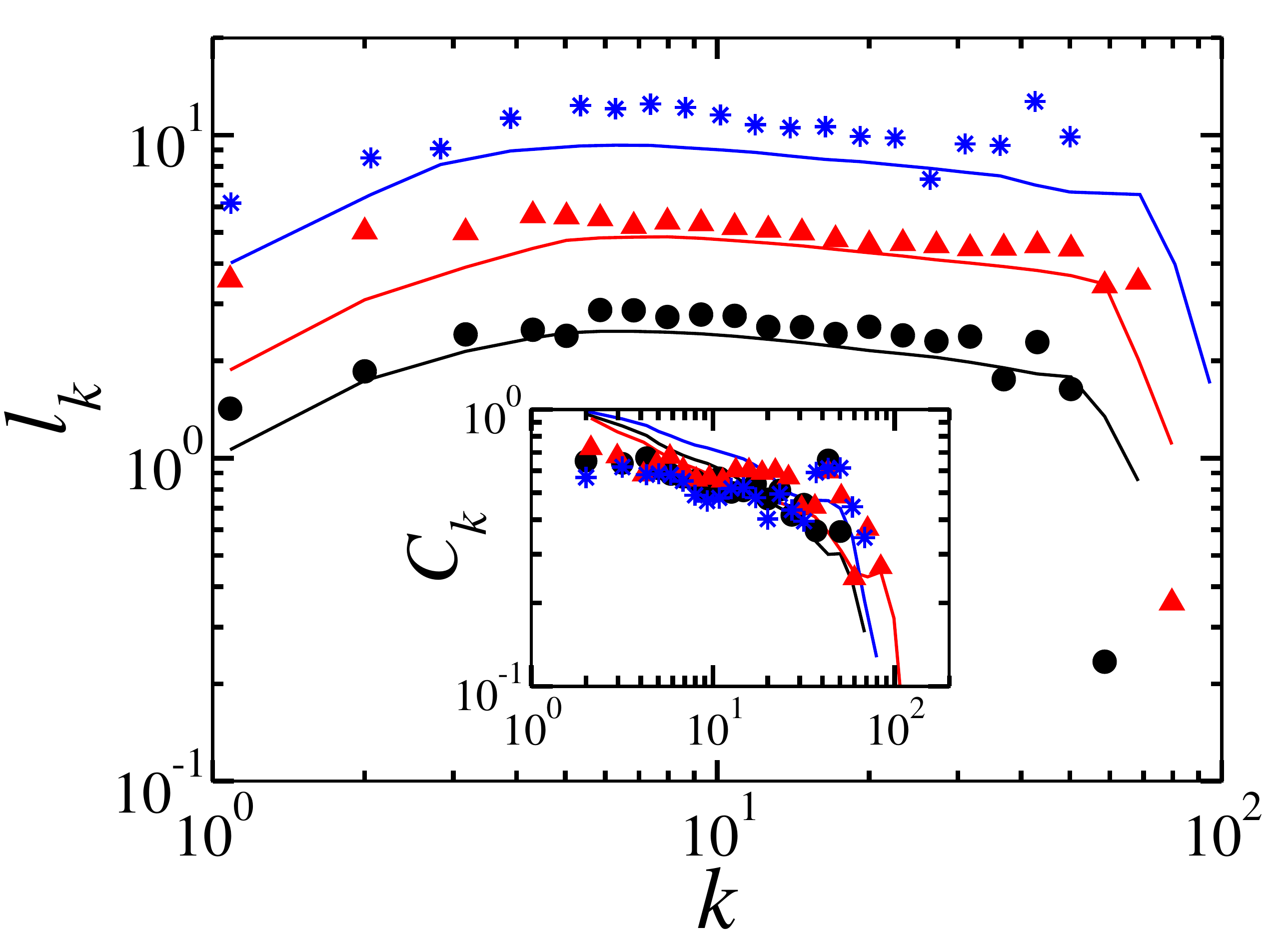}
 \caption{(Color online) Visualization of small-world properties of biological networks. Average shortest path length $l_k$ of sites of degree $k$ versus $k$ for AH (circles), EC (triangles), and VC (stars) and clustering coefficient $C_k$(inset) of sites of degree $k$ versus $k$ for AH (circles), BJ (triangles), and EC (stars), and their corresponding numerical networks (lines). Top and bottom data sets for $l_k$ are shifted vertically by factors of $2$ and $0.5$.}
\label{fig:4}
\end{figure}
In the procedure the smallest allowed degree is $k = 1$; the model then generates one large network and many small clusters, as in biological systems. We characterize this complex structure by evaluating the cluster size distribution. The cluster size is defined in terms of both the number of nodes, $S_N$, and the number of edges, $S_M$, belonging to the cluster. Figure \ref{fig:3} shows the cluster size distributions  for different biological and numerical networks. Both distributions exhibit a regime consistent with a power law with an exponent $\simeq -4.4$, for the size in terms of sites, and an exponent $\simeq -2.7$ for the size in terms of edges. The faster decay found for the first distribution suggests that the structure is highly clustered, as will be confirmed later. Furthermore, in most cases  the size of the largest connected cluster is comparable 
(Table \ref{tab2}). Interestingly, numerical data for $f(S_M)$ also reproduce the fluctuations at small sizes observed in biological data. These are not the effect of statistical noise, but measure the relative weight of the population of clusters 
with few edges, whose patterns can be simply identified. 
\begin{table}
 \begin{tabular}{c|c|c|c|c||c|c|c|c}
 \multicolumn{1}{c}{} & \multicolumn{4}{c}{\it Biological} & \multicolumn{4}{c}{\it
Model}\\
 {\it System} & $S_N^\text{max}$ & $S_M^\text{max}$ & $C_\text{max}$ &
$l_\text{max}$ & $S_N^\text{max}$ & $S_M^\text{max}$ & $C_\text{max}$ &
$l_\text{max}$ \\\hline
AH & 1785 & 7986 & 0.5 & 5.9 & 1996 & 8390 & 0.65 & 5.08\\
BJ & 2807 & 16453 & 0.5 & 6.2 & 3551 & 18043 & 0.61 & 4.85\\
CK & 2032 & 6609 & 0.5 & 8.1 & 2471 & 7398 & 0.59 & 5.81\\
EC & 2677 & 12620 & 0.5 & 6.2 & 2354 & 12250 & 0.71 & 4.92\\
NF & 1683 & 9435 & 0.5 & 6.8 & 2569 & 11257 & 0.49 & 4.33\\
PA & 2613 & 13024 & 0.5 & 7.3 & 2778 & 13263 & 0.68 & 5.09 \\
SP & 1778 & 5911 & 0.5 & 6.4 & 2484 & 7468 & 0.54 & 5.33 \\
VC & 1717 & 7726 & 0.5 & 5.6 & 1842 & 8028 & 0.60 & 4.74 \\
{SC} & 4711 & 54570 & 0.4 & 3.7 & 3351 & 53012 & 0.81 & 3.89 \\
{HS} & 10890 & 136799 & 0.4 & 3.9 & 7864 & 133576 & 0.66 & 4.15
\end{tabular}
\caption{{Properties of the largest connected cluster for the biological networks and their model counterparts: the number of nodes $S_N^\text{max}$, the number of edges $S_M^\text{max}$, the average clustering coefficient $C_\text{max}$, and the shortest path length $l_\text{max}$. The error bars are $1\%,~2\%,~4\%$ and $2\%$, respectively.}}
\label{tab2}
\end{table}
The level of connectivity in the system is measured  by the average clustering coefficient of nodes of degree $k$ and the average shortest path between nodes of degree $k$ (Fig. \ref{fig:4}). Both quantities vary smoothly with $k$ for biological and numerical data. Both the model and biological networks are highly clustered. Moreover, biological data show that the average shortest path length slowly increases with $k$ for low connectivity degrees and then reaches a fairly stable value for a wide range of $k$, in  agreement with numerical data. This result suggests that the model network reproduces not only the distribution of connectivity degrees, but also the relative position in the network of nodes with the same $k$ value. Moreover, the high value of the clustering coefficient and the small shortest path length suggest that biological and model networks have {\it small world} properties \cite{watts}. Finally the average clustering coefficient $C_\text{max}$ and the average shortest path length $l_\text{max}$ evaluated for the largest cluster show a very weak dependence on the cluster size $S_N^\text{max}$ and exhibit (Table \ref{tab2}) a good agreement between biological and model data.\\ 
A further confirmation that our model captures the structure of the network at both a global and local level is given by the evaluation of the average degree of the neighbors of a site of degree $k$, $\overline{k}_{nn}(k)$ (Fig. \ref{fig:6}). This quantity increases with the node degree as $k^{0.67 \pm 0.02}$ for biological networks, and $k^{0.61 \pm 0.01}$ for numerical data.
This scaling behavior suggests that highly connected nodes tend to be connected with each other.\\
{\color{black}Finally we notice that, for each system, topological properties are very stable with respect to changes in the fitting parameter and the calculation of the similarity. Even if the fitting value of $\alpha$ is changed by $\pm 0.25$ or the  similarity rule is modified $\lbrack$ e.g., using $p^\text{new}_{i,j} = N_c(i,j)/(k_i + k_j)$$\rbrack$, the topological properties exhibit similar behavior. It is also possible to infer the $\alpha$ value by analysis of only $75\%$ of the entire protein data set.}\\
{\color{black}From the statistical point of view our model seems to be a good candidate for modeling the topology of protein interaction networks. However, the ingredients we implement are not well established for protein interaction networks, although they are present in other biological systems. Stem cells are an example of depletion. When a stem cell specializes and becomes a particular cell (a red blood cell, a muscle cell, or even a neuron) it loses the ability to interact with cells from other types \cite{Gage}. Moreover, the similarity concept can be interpreted as the establishment of interacting protein families \cite{Park,Teichmann}.\\}
\section{Discussion}
{\color{black}In conclusion, we present a statistical model, which reproduces surprisingly well many topological properties of protein interaction networks. The model is based on a twofold mechanism for evolution, namely, preferential depletion and similarity. By fitting a single parameter, we are able to generate networks that reproduce protein interaction networks for different bacteria as well as {\it Saccharomyces cerevisae} and {\it Homo sapiens}. We wish to stress that not only do the largest clusters exhibit the same connectivity properties but also the small-cluster distributions show very good agreement between biological and model data. The clustering coefficient and the average path length suggest that highly connected nodes are placed in the largest cluster and preferentially connected to nodes with high degree. The systematic analysis of the network structure for a number of biological systems indicates that protein interaction networks are not scale-free but rather exhibit small-world properties. Further research should be performed to better understand the origin of this dual mechanism in protein interaction networks.\\
}
We acknowledge financial support from the ETH Competence Center ``Coping with Crises in Complex Socio-Economic Systems´´ (CCSS) through ETH Research Grant No. CH1-01-08-2 and FUNCAP.

\end{document}